\documentstyle[12pt]{article}

\renewcommand{\S}{{}^{(2)}\!S_{\mbox{\scriptsize vac}}}
\newcommand{\T}{T_{\mbox{\scriptsize vac}}^{\mu\nu}}
\renewcommand{\O}{{\cal O}}
\newcommand{\rr}{(\nabla r)^2}
\newcommand{\ru}{(\nabla r,\nabla u)}
\newcommand{\vu}{(\nabla v,\nabla u)}
\newcommand{\I}{{\cal I}^+}
\newcommand{\mI}{{\cal I}^-}
\renewcommand{\|}{\biggl |_{{\cal I}^+}}
\newcommand{\pin}{\psi^{\mbox{\scriptsize in}}_{la}}
\newcommand{\pout}{\psi^{\mbox{\scriptsize out}}_{lA}}
\newcommand{\pinstar}{\psi^{\dagger\,\mbox{\scriptsize in}}_{la}}
\newcommand{\poutstar}{\psi^{\dagger\,\mbox{\scriptsize out}}_{lA}}
\newcommand{\ein}{\varepsilon_{\mbox{\scriptsize in}}}
\newcommand{\eout}{\varepsilon_{\mbox{\scriptsize out}}}
\begin{document}

\begin{center}
{\LARGE\bf
Radiation equations for black holes}
\end{center}
\begin{center}
{\bf G.A. Vilkovisky}
\end{center}
\begin{center}
Lebedev Physical Institute, and Lebedev
Research Center in Physics,\\ Leninsky Prospect 53, 119991 Moscow,
Russia.\\ E-mail: vilkov@sci.lebedev.ru \\
\end{center}
\vspace{2cm}
\hspace{7cm}{\large\it In memory of Bryce DeWitt}
\vspace{2cm}
\begin{abstract}

It has been shown in the previous paper that the metric in the
semiclassical region of the collapse spacetime is expressed
purely kinematically through the Bondi charges. Here the Bondi
charges are expressed through this metric by calculating the
vacuum radiation against its background. The result is closed
equations for the metric and the Bondi charges. Notably, there
is a nonvanishing flux of the vacuum-induced matter charge.
\end{abstract}
\newpage

$$  $$

In ref. [1], a new approach is proposed to the problem of
backreaction of the Hawking radiation. For the spherically
symmetric collapse of a compact matter source it has been
shown that, in the semiclassical region [1] of the
expectation-value spacetime, the gravity equations close
purely kinematically leaving the arbitrariness only in the
data functions. The data functions are two Bondi charges
appearing as coefficients of the expansion of the metric
at the future null infinity $\I$
\begin{equation}
\rr\|=1-\frac{2{\cal M}(u)}{r}+\frac{Q^2(u)}{r^2}+
O\left(\frac{1}{r^3}\right)\;,
\end{equation}
\begin{equation}
\ru\|=-1-\frac{c_1}{r}+O\left(\frac{1}{r^2}\right)\;.
\end{equation}
Here $c_1$ is a constant [1], and, for the model of the vacuum
considered below, $c_1=0$. The notation in eqs. (1), (2) and
below is conventional for spherical symmetry and is the same
as in ref. [1] with the following exception. The retarded
time $u^+$ normalized at the future null infinity is in the
present notation just $u$, and the retarded time $u^-$ counted
out by an early falling observer [1] will here be denoted as
\begin{equation}
u^-=U(u)\;,\qquad \frac{du^-}{du^+}={\dot U}(u)\;.
\end{equation}
The advanced time $v$ that also figures below remains normalized
at the past null infinity $\mI$. The partial derivative
$\partial_u$ is the derivative at fixed $v$.

As shown in ref. [1], the metric in the semiclassical region
depends only on the Bondi charges ${\cal M}(u)$ and $Q^2(u)$.
On the other hand, the Bondi charges depend only on the metric
in the semiclassical region, this dependence being already
a subject of the quantum dynamics. Having both dependences
obtained, one gets self-consistent equations for the Bondi
charges and, thereby, for the expectation value of the metric
in the semiclassical region.

The calculation thus falls into three stages. The first stage:
solving the kinematical equations for the metric in terms
of the Bondi charges is accomplished in ref. [1]. The purpose
of the present work is the second stage: calculation of the
vacuum radiation against the thus obtained gravitational
background. This should express the Bondi charges through
themselves, and there will remain only the last stage:
solving the resultant self-consistent equations.

Both Bondi charges ${\cal M}(u)$ and $Q^2(u)$ appear in the
expansion of the flux component of the energy-momentum
tensor at $\I$:
\begin{equation}
\int d^2{\cal S}\,r^2\,T^{\mu\nu}
\left.\frac{\nabla_{\mu}v\nabla_{\nu}v}{\vu^2}\right|_{\I}
=-\partial_u{\cal M}+\frac{1}{2}\partial_uQ^2\frac{1}{r}
+O\left(\frac{1}{r^2}\right)\;.
\end{equation}
Here the integration is over the unit 2-sphere ${\cal S}$
entering the product $\I=(u\mbox{-axis})\times{\cal S}$.
Since the collapsing matter source is assumed having a compact
spatial support [1], the $T^{\mu\nu}$ in eq. (4) is the
energy-momentum tensor of the in-vacuum, $\T$. The quantity
(4) with $T^{\mu\nu}=\T$ is the only one that needs to be 
calculated. There is more than one way to do this calculation
in semiclassical theory. The present calculation uses the WKB
technique along the lines of refs. [2,3].

To model the vacuum, the simplest quantum field will be chosen:
a massless scalar field satisfying the equation
\begin{equation}
(\Box-\xi R){\bf\Phi}=0\;,\qquad\xi=\mbox{const.}\;,
\end{equation}
and having the energy-momentum tensor
\begin{eqnarray}
{\bf T}^{\mu\nu}&\!=\!&
(1-2\xi)\nabla^{\mu}{\bf\Phi}\nabla^{\nu}{\bf\Phi}
-2\xi{\bf\Phi}\nabla^{\mu}\nabla^{\nu}{\bf\Phi}
+\xi R^{\mu\nu}{\bf\Phi}^2\nonumber\\
&&{}+g^{\mu\nu}\left[(2\xi-\frac{1}{2})(\nabla{\bf\Phi})^2
+2\xi{\bf\Phi}\Box{\bf\Phi}-\frac{1}{2}\xi R{\bf\Phi}^2\right]\;.
\end{eqnarray}
The calculation will be carried out with an arbitrary $\xi$ but
the case of interest is $\xi=1/6$. The latter is important because
the locality of the trace of $\T$ is assumed in ref. [1].

The $\T$ at $\I$ can be obtained by a direct averaging of the
operator (6):
\begin{equation}
\T\|=\Bigl[\langle\mbox{in vac}|{\bf T}^{\mu\nu}|\mbox{in vac}\rangle
-\langle\mbox{out vac}|{\bf T}^{\mu\nu}|\mbox{out vac}\rangle
\Bigr]\|\;,
\end{equation}
and here one of the main points emerges. The subtraction in eq. (7)
is necessary because there is a noise at $\I$ even when the 
background field is absent. The principal requirement that the 
subtraction term must satisfy is its locality in the background
field. In the effective-action technique, this requirement is
secured by the locality of the counterterms. In the present
technique, it should be secured by the choice of the quantum
state in the subtraction term. The definition of this state
should not involve the quantum-field operators in the future
or past of the observation point of $\T$. With the observation
point at $\I$, the normalization state is the out-vacuum.
Below, the notation is introduced
\begin{equation}
\langle{\bf X}\rangle=
\langle\mbox{in vac}|{\bf X}|\mbox{in vac}\rangle-
\langle\mbox{out vac}|{\bf X}|\mbox{out vac}\rangle\;.
\end{equation}
From eq. (6) one obtains
\begin{equation}
\T\left.\frac{\nabla_{\mu}v\nabla_{\nu}v}{\vu^2}\right|_{\I}=
\langle\,(\partial_u{\bf\Phi})^2\rangle-
\xi\,\partial^2_{uu}\langle\,{\bf\Phi}^2\rangle+
O\left(\frac{1}{r^4}\right)\;.
\end{equation}

Expanding ${\bf\Phi}$ in spherical harmonics, one can replace
the quantum field in 4 dimensions with a sequence of quantum
fields ${\bf\Phi}_l$ in 2 dimensions. The fields
${\bf\Phi}_l$, $l=0,1,\ldots$, are defined on the Lorentzian 
subspace of a spherically symmetric spacetime and, there,
satisfy the equation
\begin{equation}
{\bf\Phi}_l=\frac{1}{r}\,{\bf\Psi}_l\;,
\end{equation}
\begin{equation}
\left(\triangle-\frac{\mathop{\triangle}r}{r}
-\xi R -\frac{l(l+1)}{r^2}\right){\bf\Psi}_l=0
\end{equation}
with $\triangle$ the D'Alembert operator in this subspace. In terms
of ${\bf\Psi}_l$ one obtains from eq.~(9)
\begin{eqnarray}
\int d^2{\cal S}\,r^2\,\T\left.
\frac{\nabla_{\mu}v\nabla_{\nu}v}{\vu^2}\right|_{\I}=
\sum_{l=0}^{\infty}(2l+1)\left[
\vphantom{\left(\frac{1}{2}\right)\left(\frac{1}{r^2}\right)}
\langle\,(\partial_u{\bf\Psi}_l)^2\rangle
-\xi\,\partial^2_{uu}\langle\,{\bf\Psi}_l{}^2\rangle\right.
\qquad\qquad\quad{}\nonumber\\
{}+\left.\frac{1}{r}\left(\frac{1}{2}-2\xi\right)
\partial_u\langle\,{\bf\Psi}_l{}^2\rangle
+O\left(\frac{1}{r^2}\right)\right]\;.
\end{eqnarray}
The expansion of ${\bf\Psi}_l$ at $\I$ is readily obtained from 
eq. (11). One needs the retarded solution which is
\begin{equation}
{\bf\Psi}_l\|={\bf\Psi}_l(u)+\frac{1}{r}\,\frac{l(l+1)}{2}
\int\limits^u_{-\infty}du\,{\bf\Psi}_l(u)+
O\left(\frac{1}{r^2}\right)\;.
\end{equation}
Here ${\bf\Psi}_l(u)$ are the data for ${\bf\Psi}_l$ at $\I$.
Inserting this expansion in eq. (12) and using eq. (4) one
obtains finally
\begin{eqnarray}
-\partial_u{\cal M}&\!=\!&\sum_{l=0}^{\infty}(2l+1)\biggl[
\langle\,\Bigl(\partial_u{\bf\Psi}_l(u)\Bigr)^2\rangle
-\xi\,\partial^2_{uu}\langle{\bf\Psi}_l{}^2(u)\rangle\biggr]\;,\\
\partial_u Q^2&\!=\!&\sum_{l=0}^{\infty}(2l+1)\biggl[
(1-4\xi)\,\partial_u\langle{\bf\Psi}_l{}^2(u)\rangle
+l(l+1)\,\partial_u\langle{\bf\Psi}_l{}^2(u)\rangle\nonumber\\
&&\qquad\qquad\qquad\quad
{}-\xi\,l(l+1)\,\partial^3_{uuu}\Bigl\langle\,
\left(\int\limits^u_{-\infty}du\,{\bf\Psi}_l(u)\right)^2
\Bigr\rangle\biggr]\;.
\end{eqnarray}
The flux of $Q^2$ in eq. (15) appears as a total derivative
(candidate for discarding) but appearances may deceive.

Eqs. (14), (15) leave one with the expectation values of 
the form
\begin{equation}
\langle\,\Bigl({\cal D}\,{\bf\Psi}_l(u)\Bigr)^2\rangle=
\langle\,\Bigl({\cal D}\,{\bf\Psi}_l(x)\Bigr)^2\rangle
\biggl|_{x\in\I}
\end{equation}
where ${\cal D}={\cal D}(\partial_u)$ is some (retarded)
linear operator. Let $\pin(x)$, $\,a=\ein,\,$ be the solution
of eq. (11) that asymptotically at $\mI$ becomes the
eigenfunction of the energy operator with the eigenvalue
$\ein$. Let $\pout(x)$, $\,A=\eout,\,$ be the solution defined
by a similar condition at $\I$. Normalized with the aid of
the inner product, the $\pin(x)$ and $\pout(x)$ make two
bases of solutions of eq. (11) (for each $l$), related by
the Bogolyubov transformation
\begin{equation}
\pout(x)=\alpha^l_{Aa}\,\pin(x)+\beta^l_{Aa}\,\pinstar(x)\;.
\end{equation}
(Summation is assumed over the repeating $a$ or/and $A$ but not $l$.)
The dagger designates complex conjugation, and "the complex conjugate
quantity" will be abbreviated as~c.c.
Expanding the quantum field ${\bf\Psi}_l(x)$ in any of the
bases and calculating the difference in eq. (8) one obtains
\begin{equation}
\langle\,\Bigl({\cal D}\,{\bf\Psi}_l(x)\Bigr)^2\rangle=
-\Bigl({\cal D}\,\poutstar(x)\Bigr)
\Bigl({\cal D}\,\pinstar(x)\Bigr)\beta^l_{Aa}+\mbox{c.c.}\;,
\end{equation}
\begin{equation}
\beta^l_{Aa}={\rm i}\int d^2x\,\,{}^{(2)}\!g^{1/2}\delta(\sigma)
(\nabla_{\mu}\sigma)\Bigl(\pin\mathop{\nabla^{\mu}}\pout(x)
-\pout\mathop{\nabla^{\mu}}\pin(x)\Bigr)\;.
\end{equation}
Here $d^2x\,\,{}^{(2)}\!g^{1/2}$ is the volume element of the 
Lorentzian subspace, and the equation
\begin{equation}
\Sigma\,\colon\qquad\sigma(x)=0\hphantom{\Sigma\,\colon\qquad}{}
\end{equation}
with $\nabla\sigma$ past-directed describes an arbitrary
Cauchy surface $\Sigma$ in this subspace.

To see that the expectation value (18) with $x\in\I$ is causal,
i.e., does not contain the background field in the future
of the observation point, note that it can be expressed
through the commutator function of the field ${\bf\Psi}_l$
\begin{equation}
G_l(x,x')=\frac{1}{{\rm i}}[{\bf\Psi}_l(x),{\bf\Psi}_l(x')]
=-{\rm i}\pin(x)\pinstar(x')+\mbox{c.c.}
=-{\rm i}\pout(x)\poutstar(x')+\mbox{c.c.}\;.
\end{equation}
Namely,
\begin{eqnarray}
\langle\,\Bigl({\cal D}\,{\bf\Psi}_l(x)\Bigr)^2\rangle
=\int d^2x'\,\,{}^{(2)}\!g'^{1/2}\delta(\sigma')
(\nabla'_{\mu}\sigma')\Bigl[\Bigl(
{\cal D}\psi^{\dagger\,\mbox{\scriptsize out}}_{lA}\,
\psi'^{\,\mbox{\scriptsize out}}_{lA}
-{\cal D}\psi^{\dagger\,\mbox{\scriptsize in}}_{la}\,
\psi'^{\,\mbox{\scriptsize in}}_{la}
\Bigr){\cal D}\mathop{\nabla'^{\mu}}G_l(x,x')\nonumber\\
{}+\Bigl(
{\cal D}\psi^{\mbox{\scriptsize in}}_{la}\mathop{\nabla'^{\mu}}
\psi'^{\dagger\,\mbox{\scriptsize in}}_{la}
-{\cal D}\psi^{\mbox{\scriptsize out}}_{lA}\mathop{\nabla'^{\mu}}
\psi'^{\dagger\,\mbox{\scriptsize out}}_{lA}
\Bigr){\cal D}G_l(x,x')\Bigr]\;.
\end{eqnarray}
With the observation point $x$ at $\I$ at a given value 
of $u$, choose the Cauchy surface $\Sigma$ as shown in Fig.1,
$\Sigma=\Sigma_1\bigcup\Sigma_2$. The contribution of
$\Sigma_2$ drops out of (22) because the commutator function
vanishes when $x'$ is outside the light cone of $x$. Then
shift $\Sigma_1$ to $\I$. The result is
\begin{eqnarray}
\langle\,\Bigl({\cal D}\,{\bf\Psi}_l(u)\Bigr)^2\rangle
=-{\rm i}\Bigl({\cal D}\,\poutstar(u)\Bigr)
\Bigl({\cal D}\,\pinstar(u)\Bigr)
\int\limits^{u+0}_{-\infty}du'
\Bigl[\pout(u')\mathop{\partial_{u'}}\pin(u')\nonumber\\
{}-\pin(u')\mathop{\partial_{u'}}\pout(u')\Bigr]+\mbox{c.c.}\;.
\end{eqnarray}
Here $\pout(u)$ and $\pin(u)$ are the data at $\I$ for the
basis functions $\pout(x)$ and $\pin(x)$. Only these data and 
only in the past of the observation point $u$ are needed to
calculate the expectation value (23).

The WKB solution for $\pin(x)$ boils down to the solution for 
the geodesics in the background metric. The background metric
is the sought for expectation value of the metric --- the final
goal of the calculation --- but kinematics gives it in terms of
arbitrary Bondi charges [1]. One needs to solve for the 
congruence of the null geodesics that start at $\mI$ with one
and the same energy $\ein$ and one and the same angular
momentum $L=\hbar l$. It suffices to know which of these
geodesics come to the point $u$ of $\I$, and what then
is their energy with respect to the Killing vector at $\I$.
Denote this energy as
\begin{equation}
\varepsilon_+=\varepsilon_+(u,\ein,L)\;.
\end{equation}
One can put the question differently. Consider the geodesic
that comes to the point $u$ of $\I$ with the energy
$\varepsilon_+$ and angular momentum $L$, and trace it back
to $\mI$. Then what is its energy with respect to the Killing
vector at $\mI$? The answer is given by eq. (24) solved with
respect to $\ein$. The result is
\begin{equation}
\varepsilon_+=\left\{
\begin{array}{ll}
{\displaystyle\ein\;,}&{\displaystyle\qquad L>\ein H(u)\;,}\\
{\displaystyle\ein\frac{dU(u)}{du\hphantom{U)(}}\;,}&
{\displaystyle\qquad L\le\ein H(u)
\frac{dU(u)}{du\hphantom{U)(}}\;.}\\
\end{array}
\right.
\end{equation}
Two functions of the background metric govern this behaviour.
One is the $U(u)$ of eq. (3), and the other one, $H(u)$, may
be interpreted as defining the variable height of the
centrifugal barrier. Both are expressed through the Bondi charges:
\begin{equation}
H^{-2}(u)=2\,\frac{{\cal M}(u)
+\sqrt{9{\cal M}^2(u)-8Q^2(u)}}{\left(3{\cal M}(u)
+\sqrt{9{\cal M}^2(u)-8Q^2(u)}\right)^3}\;,
\end{equation}
and [1]
\begin{equation}
\frac{d\hphantom{u}}{du}\ln\frac{dU(u)}{du\hphantom{U)(}}
=-\kappa(u)\;,
\end{equation}
\begin{equation}
{}\hphantom{u>u_0\;.\qquad}\kappa(u)=
\frac{\sqrt{{\cal M}^2(u)-Q^2(u)}}{\left({\cal M}(u)
+\sqrt{{\cal M}^2(u)-Q^2(u)}\right)^2}\;,\qquad u>u_0\;.
\end{equation}
Here $u_0$ labels the radial light ray tangent to the apparent
horizon [1]. Expression (28) is invalid for early $u$, and
the range of $u$ for which eq. (25) holds is $u>u_0+O({\cal M})$
but this is the range in which one needs to calculate the
radiation. At earlier $u$, the radiation is negligible. The
geodesics whose $\ein$ and $L$ do not satisfy either of the
inequalities in eq. (25) do not come to the point $u$ of $\I$.

The derivation of the result above will here be omitted. It will
only be mentioned that the geodesics corresponding to the second 
line of eq. (25) turn (in $r$) three times. The respective
particles start at $\mI$ as incoming and turn the first time
before the black hole has formed. They are outgoing already
when the collapsing mass comes and turns them back. They cross
the apparent horizon and continue falling down but, before
they reach small $r$ and even before they cross a possible
event horizon [1], gravity weakens and lets them go. They
cross the apparent horizon the second time, next turn the
third time, and go out to $\I$. The third turn is a quantum
effect.

The congruence of geodesics considered is hypersurface 
orthogonal, and the function (24) defines the phase of
$\pin(x)$ at $\I$. Hence one obtains the data at $\I$ for
the basis functions:
\begin{eqnarray}
\pin(u)&\!=\!&\frac{1}{\sqrt{4\pi\ein}}\Bigl[
\theta(\ein H{\dot U}-l)\exp(-{\rm i}\ein U)
+\theta(l-\ein H)\exp(-{\rm i}\ein u)\Bigr]\;,\\
\pout(u)&\!=\!&\frac{1}{\sqrt{4\pi\eout}}
\exp(-{\rm i}\eout u)\;.
\end{eqnarray}
The expectation values (23) will thus be expressed entirely 
through the functions $H(u)$ and $U(u)$. It is important
that, even kinematically, the Bondi charges are not completely 
arbitrary [1]. By their properties, $H(u)$ and $\kappa(u)$
in eq. (27) are macroscopic quantities whereas the derivatives
of these functions in $u$ are microscopic quantities. In the
notation of ref. [1],
\begin{equation}
|\O|<\;H,\;\frac{1}{\kappa}\;<\frac{1}{|\O|}\;,\qquad
\frac{d\hphantom{u}}{du}H=\O\;,\qquad
\frac{d\hphantom{u}}{du}\frac{1}{\kappa}=\O\;,
\end{equation}
\begin{equation}
{}\hphantom{u>u_0\qquad}{\dot U}=\O\;,\qquad u>u_0
\end{equation}
where $\O$ vanishes when the quantum parameter tends to zero.

In the upper limit of the integral in eq. (23) one may exploit
condition (32):
\begin{equation}
{\dot U}(u+0)<{\dot U}(u)=\O
\end{equation}
and set ${\dot U}(u+0)=0$. In view of eqs. (27) and (31), this
is equivalent to setting $u+0=\infty$, and then the integration 
by parts proves that the two terms of the integral are equal. 
In the product of two functions (29) that figures in eq. (23)
only one diagonal term survives and takes the form
\begin{equation}
\pinstar(u)\,\pin(u')=\int\limits_0^{\infty}
\frac{d\ein}{4\pi\ein}\,\theta(\ein H{\dot U}-l)
\exp\Bigl({\rm i}\ein(U-U')\Bigr)
\end{equation}
because, for $u>u'$, $H{\dot U}<H'{\dot U}'$. The contribution of 
the other diagonal term to the expectation values vanishes because
the contribution of the term (34) vanishes when $\kappa\equiv0$,
and expression (23) vanishes when
$\psi^{\mbox{\scriptsize in}}=\psi^{\mbox{\scriptsize out}}$.
The cross terms vanish because they imply ${\dot U}>|\O|$ or
${\dot U}'>|\O|$.

It follows from eqs. (23) and (34) that
\begin{equation}
\sum_{l=0}^{\infty}p_1(l)
\langle\,\Bigl({\cal D}\,{\bf\Psi}_l(u)\Bigr)^2\rangle=p_1(0)
\langle\,\Bigl({\cal D}\,{\bf\Psi}_l(u)\Bigr)^2\rangle
\biggl|_{l=0}+\int\limits_0^{\infty}dl\,p_2(l)
\langle\,\Bigl({\cal D}\,{\bf\Psi}_l(u)\Bigr)^2\rangle
\end{equation}
where $p_1(l)$ is any given polynomial, and $p_2(l)$ is some
other polynomial. Doing the sum over $l$ first and the sum
over $\eout$ last, one obtains each contribution to (35)
as a spectral integral over the energy $\eout$, the spectral
function being a combination of the functions
\begin{equation}
I_n(\eout,u)=\int\limits_0^{\infty}d\ein\,{\dot U}
\left({\rm i}\ein{\dot U}\right)^n\int\limits^{\infty}_{-\infty}
du'\,\ein{\dot U}'\exp\Bigl({\rm i}(\Omega-\Omega')\Bigr)\;,
\end{equation}
\begin{equation}
\Omega-\Omega'=\ein(U-U')+\eout(u-u')\;.
\end{equation}
Specifically,
\begin{eqnarray}
\langle\,\Bigl(\partial_u{\bf\Psi}_l(u)\Bigr)^2\rangle
\biggl|_{l=0}&\!=\!&\frac{2}{(4\pi)^2}\int\limits_0^{\infty}
d\eout\,I_0(\eout,u)+\mbox{c.c.}\;,\\
\langle{\bf\Psi}_l{}^2(u)\rangle\biggl|_{l=0}
&\!=\!&\frac{2}{(4\pi)^2}
\int\limits_0^{\infty}d\eout\,\frac{1}{{\rm i}\eout}
I_{-1}(\eout,u)+\mbox{c.c.}\;.
\end{eqnarray}

In eq. (36) introduce the new integration variables
\begin{equation}
y=\ein{\dot U}\frac{1}{\kappa}\;,\qquad
x=\ein{\dot U}'\frac{1}{\kappa'}
\end{equation}
to obtain
\begin{equation}
I_n(\eout,u)=\kappa^{n+1}\int\limits_0^{\infty}dy\,({\rm i}y)^n
\int\limits_0^{\infty}\frac{dx}{w}
\exp\Bigl({\rm i}(\Omega-\Omega')\Bigr)\;,
\end{equation}
\begin{equation}
w=1-\frac{d\hphantom{u'}}{du'}\frac{1}{\kappa'}\;.
\end{equation}
Next use eq. (27) to write
\begin{equation}
{\dot U}=\exp\Bigl(-\int\limits^u_{-\infty}
d{\bar u}\,{\bar\kappa}\Bigr)\;,
\end{equation}
\begin{equation}
U'-U=\int\limits_u^{u'}du''\,
\exp\Bigl(-\int\limits^{u''}_{-\infty}
d{\bar u}\,{\bar\kappa}\Bigr)
\end{equation}
and integrate in $u''$ by parts:
\begin{eqnarray}
U'-U&\!\!=\!\!&\Biggl(1+\frac{d\hphantom{u}}{du}\frac{1}{\kappa}+
\frac{d\hphantom{u}}{du}\left(\frac{1}{\kappa}
\frac{d\hphantom{u}}{du}\frac{1}{\kappa}\right)+\cdots\Biggr)
\frac{1}{\kappa}{\dot U}\nonumber\\
&&{}-\Biggl(1+\frac{d\hphantom{u'}}{du'}\frac{1}{\kappa'}+
\frac{d\hphantom{u'}}{du'}\left(\frac{1}{\kappa'}
\frac{d\hphantom{u'}}{du'}\frac{1}{\kappa'}\right)+\cdots\Biggr)
\frac{1}{\kappa'}{\dot U}'\;.
\end{eqnarray}
Owing to condition (31), all the corrections with the derivatives
of $\kappa$ are negligible both in (42) and (45):
\begin{equation}
w=1\;,\qquad\ein(U-U')=x-y\;.
\end{equation}
There remains to be expressed through $x$ and $y$ the difference
$(u-u')$ in eq. (37). For that one has the equation
\begin{equation}
\ln\frac{y}{x}=\int\limits_u^{u'}du''\,
\kappa''\left(1-\frac{d\hphantom{u''}}{du''}
\frac{1}{\kappa''}\right)
\end{equation}
in which the last term is negligible. This equation can be solved
by expanding in the derivatives of $\kappa$:
\begin{equation}
\kappa(u'-u)=\ln\frac{y}{x}+\frac{1}{2}
\left(\frac{d\hphantom{u}}{du}\frac{1}{\kappa}\right)
\ln^2\frac{y}{x}+\frac{1}{6}\Biggl(
\frac{d\hphantom{u}}{du}\left(\frac{1}{\kappa}
\frac{d\hphantom{u}}{du}\frac{1}{\kappa}\right)\Biggr)
\ln^3\frac{y}{x}+\cdots
\end{equation}
but here the corrections with the derivatives are not
unconditionally negligible as they are in eqs. (42)
and (45). The point here is that the integrals in $x$ and $y$
are cut off by oscillations at both the upper and lower
limits so that the integration regions for $\ln x$ and
$\ln y$ are effectively compact. Then $\ln(y/x)$ is bounded,
and, in eq. (48), all the terms with the derivatives of $\kappa$
are negligible. As a consequence, the spectral function (41)
is calculable. Another important consequence is
\begin{equation}
\kappa(u'-u)<\frac{1}{|\O|}\;.
\end{equation}
It was tacitly assumed in the derivations above that conditions
(31)-(32) valid at the observation point $u$ are valid also
at the integration point $u'$. Eq. (49) proves that this is
the case.

Now one comes to the central point of the present consideration.
The argument above about the effective compactness of the
integration regions for $\ln x$ and $\ln y$ needs a reserve.
At the upper limits, the integrals in $x$ and $y$ are cut off
by oscillations always whereas at the lower limits only when
$\eout\ne0$. Therefore, the argument may break down for the
low-energy part of the spectrum: $\eout\to0$. This can be checked.
Introduce the notation
\begin{equation}
z=\frac{\eout}{\kappa(u)}\;.
\end{equation}
If the corrections in eq. (48) are small indeed, their 
contributions to the spectral function can be calculated by
expanding the exponential
\begin{equation}
\exp({\rm i}\eout(u-u'))=\left[1+P\Bigl({\rm i}z,\ln\frac{x}{y}
\Bigr)\right]\exp\Bigl({\rm i}z\ln\frac{x}{y}\Bigr)\;.
\end{equation}
Here $P$ is a series of the form
\begin{equation}
\hphantom{\qquad k\ge1,\quad s\ge k+1\;.}
P\Bigl({\rm i}z,\ln\frac{x}{y}\Bigr)=
\sum\O({\rm i}z)^k\Bigl(\ln\frac{x}{y}\Bigr)^s\;,
\qquad k\ge1,\quad s\ge k+1\;.
\end{equation}
Insertion of the expansion (51) in eq. (41) yields the following
result:
\begin{equation}
I_n(\eout,u)=\kappa^{n+1}(u)\left[1+
P\Bigl({\rm i}z,\frac{d\hphantom{{\rm i}z}}{d{\rm i}z}\Bigr)
\right]F_n(z)\;,
\end{equation}
\begin{equation}
F_n(z)={\rm e}^{-\pi z}\Gamma(n+1-{\rm i}z)\Gamma(1+{\rm i}z)
\end{equation}
where $\Gamma$ is the Euler's function. When $n\ge0$, the
function $F_n(z)$ and all its derivatives are bounded
including at $z\to0$. Then the entire contribution of $P$
in eq. (53) is $\O$ and is negligible. However, when $n<0$,
the function $F_n(z)$ behaves as $1/z$ at $z\to0$. Then,
because, in the series $P$, the power of $(d/d{\rm i}z)$
exceeds the power of $({\rm i}z)$ at least by one, the corrections
due to $P$ are {\it even more singular}, and, therefore,
at $\eout\to0$ they are not small.

Thus, for $n\ge0$ the spectral function (36) is successfully
calculated. One has
\begin{equation}
n\ge0\,\colon\qquad
I_n(\eout,u)=\kappa^{n+1}(u){\rm e}^{-\pi z}
\Gamma(n+1-{\rm i}z)\Gamma(1+{\rm i}z)
\hphantom{n\ge0\,\colon\qquad}
\end{equation}
with $z$ in eq. (50). Hence one obtains, in particular, the
expectation value (38)
\begin{equation}
\langle\,\Bigl(\partial_u{\bf\Psi}_l(u)\Bigr)^2\rangle
\biggl|_{l=0}=\frac{4}{(4\pi)^2}\int\limits_0^{\infty}
d\eout\,\frac{2\pi\eout}{{\rm e}^{2\pi z}-1}
\end{equation}
but does not obtain (39).

For $I_n$ with $n<0$ one has a difficulty. In order that the
function $\kappa(u)$ could be regarded as slowly varying,
the operator ${\cal D}$ acting on
$\psi^{\dagger\,\mbox{\scriptsize in}}(u)$ in eq. (23) should be
$\partial_u$ to the power 1 or higher. This suggests the way
of overcoming the difficulty. For ${{\cal D}=\partial_u{}^0}$
and ${{\cal D}=\partial_u{}^{-1}}$ write in eq. (23)
\begin{eqnarray}
\psi^{\dagger\,\mbox{\scriptsize in}}(u)&\!=\!&
\int\limits^u_{-\infty}d{\bar u}\,\partial_{\bar u}
{\bar\psi}^{\dagger\,\mbox{\scriptsize in}}\;,\\
\int\limits^u_{-\infty}du\,
\psi^{\dagger\,\mbox{\scriptsize in}}(u)&\!=\!&
\int\limits^u_{-\infty}d{\bar u}\,(u-{\bar u})\,\partial_{\bar u}
{\bar\psi}^{\dagger\,\mbox{\scriptsize in}}\;.
\end{eqnarray}
Alternatively, write in eq. (36)
\begin{equation}
\exp({\rm i}\ein U)=\int\limits^u_{-\infty}d{\bar u}
\left({\rm i}\ein{\dot{\overline{U}}}\right)
\exp({\rm i}\ein\overline{U})
\end{equation}
to obtain
\begin{equation}
I_{-1}(\eout,u)=\int\limits^u_{-\infty}d{\bar u}\,
I_0(\eout,{\bar u})
\exp({\rm i}\eout(u-{\bar u}))\;.
\end{equation}
This calculates $I_{-1}$ through $I_0$, and for $I_0$ one has
the result (55) but the new obstacle is that the integral (60)
involves $u$ down to $u=-\infty$ whereas the result (55) is
valid only for $u>u_0$. Indeed, only at late $u$ is $\kappa(u)$
slowly varying by virtue of condition (31). The obstacle is
not big, however. Expression (55) is inaccurate at early time
but, since the radiation at early time is negligible altogether,
this inaccuracy is unessential. The specific form of $\kappa(u)$
at early time is also unessential. It is only important that
$\kappa(u)$ falls off at $u\to-\infty$ so that the integral in
eq. (43) converges. For the calculational purposes, one may
just set $\kappa(u)=0$ for $u<u_0$, this being equivalent to
switching off the background curvature at early time.

Using eqs. (60) and (55), one obtains the expectation value (39):
\begin{equation}
\langle{\bf\Psi}_l{}^2(u)\rangle\biggl|_{l=0}=
\frac{4}{(4\pi)^2}\int\limits_0^{\infty}d\eout
\int\limits^u_{-\infty}d{\bar u}\,
\frac{2\pi\sin\left(\eout(u-{\bar u})\right)}{{\rm e}^{2\pi{\bar z}}
-1}
\end{equation}
where ${\bar z}=\eout/{\bar\kappa}$, and
${\bar\kappa}=\kappa({\bar u})$.
The integral over $\eout$ can be done:
\begin{equation}
\langle{\bf\Psi}_l{}^2(u)\rangle\biggl|_{l=0}=
\frac{4}{(4\pi)^2}\int\limits^u_{-\infty}d{\bar u}\,{\bar\kappa}\,
\left[\frac{\pi}{2}-\frac{\pi}{{\bar\kappa}(u-{\bar u})}+
\frac{\pi}{{\rm e}^{{\bar\kappa}(u-{\bar u})}-1}\right]\;,
\end{equation}
and finally one obtains
\begin{equation}
\langle{\bf\Psi}_l{}^2(u)\rangle\biggl|_{l=0}=
\frac{4}{(4\pi)^2}\left[
\frac{\pi}{2}\Bigl(\int\limits^u_{-\infty}d{\bar u}\,
{\bar\kappa}\Bigr)
-\pi\ln\Bigl(\int\limits^u_{-\infty}d{\bar u}\,{\bar\kappa}\Bigr)
+O(1)\right]\;,\qquad
\Bigl(\int\limits^u_{-\infty}d{\bar u}\,{\bar\kappa}\Bigr)\to\infty\;.
\end{equation}
This is the message of the present paper. The operator
${\bf\Phi}^2\Bigl|_{\I}$ averaged over the in-vacuum or 
out-vacuum is infrared-divergent. The expectation value
$\langle\,{\bf\Phi}^2\rangle\Bigl|_{\I}$
obtained as the difference in eq. (8) is finite but {\it growing
at late time}. Its first derivative contributes significantly
to the fluxes (14)-(15):
\begin{equation}
\partial_u\langle{\bf\Psi}_l{}^2(u)\rangle\biggl|_{l=0}=
\frac{4}{(4\pi)^2}\,\frac{\pi}{2}\kappa(u)(1+\O)
\end{equation}
whereas the second derivative is already negligible:
\begin{equation}
\partial^2_{uu}\langle{\bf\Psi}_l{}^2(u)\rangle\biggl|_{l=0}=
O\Bigl(\frac{d\kappa}{du}\Bigr)=\kappa^2\O\;.
\end{equation}
This mechanism of emergence of the vacuum fluxes is familiar.
In the effective-action technique they {\it all} emerge as
total derivatives of growing vertex functions [4].

An alternative way of calculating expression (61) is introducing
the integration variables
\begin{equation}
\hphantom{\qquad\quad{\bar u}>u_0}
\gamma=\eout(u-{\bar u})\;,\qquad
\sigma={\bar\kappa}(u-{\bar u})\;,\qquad\quad
{\bar u}>u_0
\end{equation}
and doing the integral over $\sigma$ first. It works also when
calculating the $l>0$ contributions to the expectation values.
Using eqs. (57), (58) one obtains
\par\smallskip
\begin{eqnarray}
\makebox[14cm][l]{$\displaystyle
\int\limits_0^{\infty}dl\,l^n\,
\langle{\bf\Psi}_l{}^2(u)\rangle=
\frac{4}{(4\pi)^2}\frac{n!}{n+1}H^{n+1}(u)\kappa^{n+1}(u)$}\nonumber\\
{}\times\left\{
\begin{array}{ll}
{\displaystyle -(-1)^{\frac{n}{2}}\ln\Bigl(
\int\limits^u_{-\infty}d{\bar u}\,{\bar\kappa}\Bigr)+O(1)\;,}&
{\displaystyle\quad\; n\mbox{ even}}\\
{\displaystyle O(1)\;,}&{\displaystyle\quad\; n\mbox{ odd}\;,}
\end{array}
\right.
\end{eqnarray}
\begin{eqnarray}
\makebox[14cm][l]{$\displaystyle
\int\limits_0^{\infty}dl\,l^n\,
\Bigl\langle\,\left(\int\limits^u_{-\infty}du\,
{\bf\Psi}_l(u)\right)^2\Bigr\rangle=
\frac{1}{(4\pi)^2}\frac{n!}{n+1}H^{n+1}(u)\kappa^{n-1}(u)$}\nonumber\\
{}\times\left\{
\begin{array}{ll}
{\displaystyle (-1)^{\frac{n}{2}}\Bigl(
\int\limits^u_{-\infty}d{\bar u}\,{\bar\kappa}\Bigr)^2+
O\Bigl(\int\limits^u_{-\infty}d{\bar u}\,{\bar\kappa}\Bigr)\;,}&
{\displaystyle\quad\; n\mbox{ even}}\\
{\displaystyle (-1)^{\frac{n+1}{2}}2\pi\Bigl(
\int\limits^u_{-\infty}d{\bar u}\,{\bar\kappa}\Bigr)+
O\Bigl(\ln\Bigl(
\int\limits^u_{-\infty}d{\bar u}\,{\bar\kappa}\Bigr)\Bigr)\;,}&
{\displaystyle\quad\; n\mbox{ odd}\;.}
\end{array}
\right. 
\end{eqnarray}
Here the powers of growth are different for even and odd powers
of $l$ but in all cases this growth is insufficient. Since, in
eq. (15), on the terms (67) there acts $\partial_u$ and on the terms
(68) $\partial^3_{uuu}$, the contributions of all of these
terms are negligible. {\it Only the s-mode contributes to the
flux of the charge} $Q^2$.

The $l>0$ modes contribute only to the flux of the gravitational
charge ${\cal M}$ via
\begin{eqnarray}
\makebox[14cm][l]{$\displaystyle
\int\limits_0^{\infty}dl\,l^n\,
\langle\,\Bigl(\partial_u{\bf\Psi}_l(u)\Bigr)^2\rangle=
\frac{2}{(4\pi)^2}\frac{1}{n+1}H^{n+1}(u)\kappa^{n+1}(u)$}\nonumber\\
{}\times\int\limits_0^{\infty}d\eout\,\eout\left[
-(-{\rm i})^n{\rm e}^{-\pi z}\Gamma(n+1-{\rm i}z)
\Gamma(1+{\rm i}z)+\mbox{c.c.}\right]\;.
\end{eqnarray}
The WKB technique is inaccurate for $l$ of order 1 but the
contribution of the s-mode is unambiguous and so is the
contribution of the $l\gg1$ modes. The latter is given by the
highest power of $l$ in the polynomial $p_2(l)$ in eq. (35).
Retaining only these two contributions one obtains
\begin{equation}
-\partial_u{\cal M}=\frac{4}{(4\pi)^2}2\pi
\int\limits_0^{\infty}d\eout\,
\frac{\eout+H^2(u)
\varepsilon^3_{\mbox{\scriptsize out}}}{{\rm e}^{2\pi z}-1}\;.
\end{equation}
The contribution of the s-mode describes correctly the
low-energy behaviour of the spectral function, and the
contribution of the $l\gg1$ modes describes correctly the
high-energy behaviour. The inaccuracy at intermediate energies
is a question of the grey-body factor.

The final result is the following set of equations for the
Bondi charges:
\begin{eqnarray}
-\partial_u{\cal M}&\!=\!&\frac{1}{48\pi}\,\kappa^2(u)
\left(1+\frac{1}{10}\left(H(u)\kappa(u)\right)^2\right)\;,\\
\partial_uQ^2&\!=\!&\frac{1}{8\pi}\,\kappa(u)(1-4\xi)\;.
\end{eqnarray}
The contribution of the $l\gg1$ modes to the total energy
flux (71) is by an order of magnitude less than the
contribution of the s-mode because the high-energy part
of the Planckian spectrum is exponentially suppressed.
The flux of the charge $Q^2$ depends on the value of $\xi$,
and there are values for which it is zero or negative but,
if $\xi\ne1/6$, the present calculation is inconsistent
because the background metric of ref. [1] is invalid. The
nonvanishing flux of the charge $Q^2$ is a surprise.

In conclusion, a failure of the 2-dimensional effective action
is worth mentioning. At $\xi=0$, to the s-mode ${\bf\Psi}_0$
there corresponds the effective action in the Lorentzian subspace:
\begin{equation}
\S=-\frac{1}{96\pi}\int d^2x\,\,{}^{(2)}\!g^{1/2}\,\,
{}^{(2)}\!R\frac{1}{\triangle}{}^{(2)}\!R
\end{equation}
so that
\begin{equation}
\frac{2}{{}^{(2)}\!g^{1/2}}\,
\frac{\delta\,\,\S}{\delta g_{\mu\nu}\hphantom{\,\,{}^{(2)}}}\,
\frac{\nabla_{\mu}v\nabla_{\nu}v}{\vu^2}=
\langle\,\Bigl(\partial_u{\bf\Psi}_0\Bigr)^2\rangle
\end{equation}
(with the retarded current on the left-hand side). Eq. (74) is to be
compared with eq. (12) at $\xi=0$ and $l=0$. Only to the leading
order in $1/r$ do these expressions coincide. The effective action
(73) should, therefore, give the correct energy flux for the
s-mode, and it does. But the flux of the charge $Q^2$ is not
contained in this action even for the s-mode and even at $\xi=0$.
This may explain why the 2-dimensional models of the effective
equations miss the backreaction of radiation. The 4-dimensional
effective action should, of course, reproduce the present results in
full but here it will not be considered.

The equations above for the Bondi charges close. Thereby, the
expectation-value equations for the metric close already at
the level of functions of one variable [1]. The solution will be
reported.

The present work was supported by the Italian Ministry for
Foreign Affairs via Centro Volta, and the Ministry of Education
of Japan via the Yukawa Institute for Theoretical Physics.
Special thanks to Luigi Cappiello and Roberto Pettorino,
and Masao Ninomiya and Mihoko Nojiri for their hospitality in
Naples and Kyoto respectively.

\newpage

\begin{center}
\section*{\bf References}
\end{center}

$$ $$

\begin{enumerate}
\item G.A. Vilkovisky, {\it Kinematics of evaporating black holes},
the accompanying paper.
\item S.W. Hawking, Commun. math. Phys. 43 (1975) 199.
\item B.S. DeWitt, Phys. Rep. 19 (1975) 295.
\item A.G. Mirzabekian and G.A. Vilkovisky, Phys. Lett. B 414
(1997) 123; Ann. Phys. 270 (1998) 391.
\end{enumerate}

\newpage

\begin{center}
\section*{\bf Figure caption}
\end{center}

$$ $$

\begin{itemize}
\item[Fig.1.] Choice of the Cauchy surface $\Sigma$ for the
inner product, $\Sigma=\Sigma_1\bigcup\Sigma_2$. The division
into $\Sigma_1$ and $\Sigma_2$ is specified by the location
of the observation point $u$ at $\I$. The event horizon EH
if any is in the future of $\Sigma$.
\end{itemize}

\end{document}